\documentclass[preprint,showpacs,preprintnumbers,amsmath,amssymb]{revtex4}

\begin{document}
%\newpage
\title{
Ising model with mixed boundary conditions: \\
universal amplitude
ratios. }

\author{N. Sh. Izmailian,$^{1,2,3}$  and
Yeong-Nan Yeh$^{1}$}

\affiliation{$^1$ Institute of Mathematics, Academia Sinica,
Nankang, Taipei 11529, Taiwan, R.O.C.}

\affiliation{$^2$Yerevan Physics Institute, Alikhanian Br. 2,
375036 Yerevan, Armenia}

\affiliation{$^3$ International Center for Advanced Study, Yerevan
State University, 1 Alex Manoogian St., Yerevan, 375025, Armenia}

\date{\today}

\begin{abstract}
In the vicinity of boundaries the bulk universality class of critical phenomena splits into several boundary universality classes, depending upon
whether the tendency to order in the boundary is smaller or
larger than in the bulk. For Ising universality class there are five different boundary universality classes: periodic, antiperiodic, free, fixed and mixed (mixture of the last two). In this paper we present the new set of the universal amplitude ratios for the mixed boundary universality class. The results are in perfect agreement with a perturbated conformal field theory scenario
proposed by Cardy \cite{cardy86}.
\end{abstract}

\maketitle

\vskip 1cm
%%%%%%%%%%%%%%%%%%%%%%%%
%\section{Introduction}
%\label{introduction}
%%%%%%%%%%%%%%%%%%%%%%%%
A central element of the modern theory of bulk critical phenomena is the division into (bulk) universality classes. As is well known, the critical behavior near boundaries normally differs from the bulk behavior. In general, each bulk
universality class of critical phenomena splits into several surface universality classes.

The criteria by which the critical systems can be classified into
different universality classes is a problem of much academic
interest. Two-dimensional critical systems are parameterized by
the conformal anomaly $c$ which is the central charge in the
Virasoro algebra \cite{Belavin,Dotsenko}. The conformal anomaly
$c$ can be obtained directly from the finite size corrections to
the free energy for a two-dimensional classical system on
infinitely long but finitely wide strip at a conformally invariant
critical point.

The asymptotic finite-size scaling behavior of the critical free
energy $f_{\cal M}$ and the critical inverse correlation length
$\xi_i$ associated with the two-point correlations of in a two-dimensional spin system  on a strip
with infinite length and a width of ${\cal M}$ lattice spacings
has the form
\begin{equation}
\lim_{{\cal M} \to \infty} {{\cal M}^2 (f_{\cal M}-f_{\infty})- 2 {\cal M} f_{surf}}=A,
\label{I2}
\end{equation}
\begin{equation}
\lim_{{\cal M} \to \infty} {{\cal M}
\xi_{i}^{-1}}=D_{i}, \label{I1}
\end{equation}
where $f_{\infty}$ is the bulk free energy, $f_{surf}$ is the
surface free energy and $A$ and $D_{i}$  are the universal
constants, but may depend on the boundary conditions (BCs).
The index $i$
distinguishes different correlation lengths: for example, $i = s$ for the spin-spin correlation
length or $i = e$ for the energy-energy correlation length.
In some two-dimensional geometries, the values of $A$ and
$D_{i}$ is known \cite{Blote,cardy86a}, to be related to the
conformal anomaly number ($c$), the highest conformal weight
$(\Delta)$, and the scaling dimensions of the $i$-th scaling field ($x_{i}$)  of the theory
\begin{eqnarray}
A&=& -4\pi \left(\Delta-\frac{c}{24}\right), \qquad D_{i} =
2\pi x_{i} \qquad \mbox{for periodic or antiperiodic BCs},
\label{Aperiod} \\
A&=& -\pi  \left(\Delta-\frac{c}{24}\right),  \qquad \quad
D_{i} = \pi x_{i} \qquad \mbox{for free, fixed and mixed
BCs}, \label{Afree}
\end{eqnarray}
The principle of unitarity of the underlying field
theory restricts through the Kac formula the possible values
of c and for each value of c only permits a finite number of
possible values of $\Delta$.
For the 2D Ising model, we have $c=1/2$ and the only possible values
are $\Delta= 0, 1/16, 1/2$.
The highest conformal weight $\Delta$, and the scaling dimension
$x_{i}$ depends on the BCs. For the Ising model on infinitely
long cylinder there are two different boundary universal classes:
periodic and antiperiodic with
\begin{eqnarray}
\Delta&=&0, \hspace{1cm} x_s=\frac{1}{8}, \qquad x_e=1 \qquad
\mbox{for periodic BCs}
\label{perD}, \\
\Delta&=&\frac{1}{16},\hspace{0.7cm} x_s=\frac{3}{8},\qquad x_e=1
\qquad  \mbox{for antiperiodic BCs.}
   \label{antiperD}
\end{eqnarray}
For the Ising model on infinitely long strip there are three
different boundary universal classes: free, fixed ($+-$) and mixed with
\begin{eqnarray}
\Delta&=&0, \hspace{1cm}
x_s=\frac{1}{2}, \qquad x_e=2 \qquad \mbox{for free and fixed ($++$) BCs} \label{freeD}, \\
\Delta&=& \frac{1}{2}, \hspace{1cm} x_s=2,\qquad x_e=2 \quad \quad
\mbox{ for fixed ($+-$) BCs}
\label{fixedD}, \\
\Delta&=&\frac{1}{16},\qquad x_s=1,\qquad x_e=2\qquad  \mbox{for
mixed BCs}. \label{mixedD}
\end{eqnarray}
For fixed $++$ (or $+-$) boundary conditions the spins are fixed to the same
(or opposite) values on two sides of the strip. The mixed boundary conditions corresponds to free boundary conditions on
one side of the strip,
and fixed boundary conditions on the other.

In the terminology of surface critical phenomena these three boundary universal classes: free, mixed and fixed ($+-$) correspond
to "ordinary", "special" and "extraordinary" surface critical behavior, respectively.

Quite recently, Izmailian and Hu \cite{izmailan,izmailan1} studied the
finite size correction terms for the free energy per spin and the
inverse correlation length of critical two-dimensional (2D) Ising
models on ${\cal M} \times \infty$ lattice and one-dimensional quantum spin model  with periodic, antiperiodic and free BCs.
They obtain analytic expressions for the finite-size correction coefficients $a_k$, $b_k$ and $c_k$ in the expansions
\begin{eqnarray}
{\cal M} \left(f_{{\cal M}}-f_{\infty}\right) &=&2  f_{surf}+\sum_{k=1}^{\infty}\frac{a_k}{{\cal M}^{2 k-1}},
\label{fN} \\
\xi_s^{-1}&=&\sum_{k=1}^{\infty}\frac{b_k}{{\cal M}^{2
k-1}}, \label{cli} \\
\xi_e^{-1}&=& \sum_{k=1}^{\infty}\frac{c_k}{{\cal M}^{2 k-1}}.
\label{clienergy}
\end{eqnarray}
and find that although the finite-size correction coefficients
$a_k$, $b_k$ and $c_k$ are not universal, the amplitude ratios for
the coefficients of these series are universal and given by
\begin{eqnarray}
r_s(k) &=& \frac{b_k}{a_k} = \frac{2^{2k}-1}{2^{2k-1}-1}; \hspace{3cm}  r_s(1)=3, \quad r_s(2)=\frac{15}{7}, \quad \dots \label{I4}\\
r_e(k)&=&\frac{c_k}{a_k}=\frac{4 k}{(2^{2k-1}-1)B_{2k}};
\hspace{2cm} r_e(1)=24, \quad r_e(2)=-\frac{240}{7}, \quad \dots.
\label{I5}
\end{eqnarray}
for periodic BCs, where $B_n$ is the $n$-th Bernoulli number ($B_2 = 1/6,B_4 =
-1/30, \dots$),

\begin{eqnarray}
r_s(k) &=& \frac{b_k}{a_k} = \frac{(2^{2k}-1)B_{2k}-2 k}{2^{2k-1} B_{2k}};
\hspace{1cm}  r_s(1)=-\frac{9}{2}, \quad r_s(2)=\frac{135}{8}, \quad \dots  \label{I8}\\
r_e(k) &=& \frac{c_k}{a_k} = -\frac{2 k}{B_{2k}}; \hspace{3.2cm}
r_e(1)=-12, \quad r_e(2)=120, \quad \dots.\label{I9}
\end{eqnarray}
for antiperiodic BCs,
\begin{eqnarray}
r_s(k) &=& \frac{b_k}{a_k} = \frac{4 k}{(2^{2k-1}-1)B_{2k}};
\hspace{2cm} r_s(1)=24, \quad r_s(2)=-\frac{240}{7}, \quad \dots  \label{I6}\\
r_e(k) &=& \frac{c_k}{a_k} = \frac{4 k
(3^{2k-1}+1)}{(2^{2k-1}-1)B_{2k}}; \hspace{2cm} r_e(1)=96, \quad
r_e(2)=-960, \quad \dots.
 \label{I7}
\end{eqnarray}
for free BCs.

In this paper we present exact calculations for a set of universal
amplitude ratios for the two-dimensional (2D) Ising
models on ${\cal M} \times \infty$ lattice with the special boundary conditions studied by Brascamp
and Kunz (BK) \cite{Brascamp}. They considered a lattice with $2 {\cal N}$ sites
in the $x$ direction and ${\cal M}$ sites in the $y$ direction. The boundary
conditions are periodic in the $x$  direction; in the $y$ direction,
the spins are up (+1) along  the upper border of the resulting cylinder
and have the alternative  values along the lower border of the resulting
cylinder. It was shown \cite{Izmailian3} that the asymptotic finite-size scaling behavior of the critical free
energy $f_{\cal M}$ of the Ising model on infinitely long strip with Brascamp-Kunz boundary condition has the form
\begin{equation}
\lim_{{\cal M} \to \infty} {{\cal M}^2 (f_{\cal M}-f_{\infty})- 2 {\cal M} f_{surf}}=-\frac{\pi}{24},
\label{I2new}
\end{equation}
which is consistent with the conformal field theory prediction for the mixed boundary condition (see Eqs. (\ref{I2}), (\ref{Afree}) and (\ref{mixedD})) although the mixed boundary condition and the BK boundary
condition are different on one side of the long strip.

 We obtain analytic
equations for $a_k$, $b_k$ and $c_k$ in the expansions given by
Eqs. (\ref{fN}), (\ref{cli}) and (\ref{clienergy}) and find that
universal amplitude ratios the two-dimensional (2D) Ising
models on ${\cal M} \times \infty$ lattice with mixed BCs are given by
\begin{eqnarray}
r_s(k) &=& \frac{b_k}{a_k} = -\frac{4k}{B_{2k}};
\hspace{2.5cm} r_s(1)=-24, \quad r_s(2)=240, \quad \dots  \label{new6}\\
r_e(k) &=& \frac{c_k}{a_k} =-\frac{2^{2k+1}k}{B_{2k}} ;
\hspace{2cm} r_e(1)=-48, \quad r_e(2)=1920, \quad \dots.
 \label{new7}
\end{eqnarray}
As far as we know, no previous
RG arguments, analytic calculations, or numerical studies predict
the existence of this whole set of universal amplitude ratios.

Consider an Ising ferromagnet on an ${\cal N} \times {\cal M}$ lattice.
The Hamiltonian of the system is
\begin{equation}
\beta H=-J\sum_{<ij>} s_i s_j,
\label{I51}
\end{equation}
where $\beta = (k_BT)^{-1}$, the Ising spins $s_i=\pm 1$ are located at
the sites of the lattice and the summation goes over all
nearest-neighbor pairs of the lattice. There are a few boundary conditions for which the Ising model has been solved exactly.
Among  them is the special boundary conditions studied by Brascamp
and Kunz (BK) \cite{Brascamp}.
We consider a transfer matrix acting along the ${\cal M}$ direction.
If $\Lambda_0$, $\Lambda_1$ and $\Lambda_2$ are the largest, the second-largest and the third-largest eigenvalues of the
transfer matrix, in the limit ${\cal N} \to \infty$
the free energy per spin, $f_{\cal M}=\lim_{{\cal N} \to \infty}F/2{\cal N}$,
$$
f_{\cal M}=\lim_{{\cal N} \to \infty}\frac{F}{2{\cal N}({\cal M}+1)}
$$ and the inverse
 spin-spin correlation length, $\xi_s^{-1}$, and the inverse
 energy-energy correlation length, $\xi_e^{-1}$,
are
\begin{equation}
f_{\cal M}=\frac{1}{{\cal M}+1}\ln{\Lambda_0},\quad
\xi_s^{-1}=\ln{(\Lambda_0/\Lambda_1)}
\quad \mbox{and} \quad
\xi_e^{-1}=\ln{(\Lambda_0/\Lambda_2)}.
\label{I61}
\end{equation}
where $F$ is the total free energy. The three leading eigenvalues
of the transfer matrix ($\Lambda_0$, $\Lambda_1$ and $\Lambda_2$)
can be obtained from exact expression for the partition function
of the Ising model on an $2{\cal N} \times {\cal M}$ rectangular
lattice under under Brascamp-Kunz BCs  \cite{Brascamp}:
\begin{eqnarray}
\Lambda_0 &=&C_{\mu} \exp\left\{\frac{1}{2}\sum_{m=0}^{2{\cal M}
+1}\omega_{\mu} \left(\frac{\pi m}
{2({\cal M}+1)}\right)\right\}, \label{L0}\\
\Lambda_1 &=&C_{\mu} \exp\left\{-2\omega_{\mu} \left(\frac{\pi}{2({\cal
M}+1)}\right)+\frac{1}{2}\sum_{m=0}^{2{\cal M} +1}\omega_{\mu}
\left(\frac{\pi m} {2({\cal M}+1)}\right)\right\},
\label{L1}\\
\Lambda_2 &=&C_{\mu} \exp\left\{-2\omega_{\mu} \left(\frac{\pi}{{\cal
M}+1}\right)+\frac{1}{2}\sum_{m=0}^{2{\cal M}+1} \omega_{\mu}
\left(\frac{\pi m}{2({\cal M}+1)}\right)\right\}, \label{L2}
\end{eqnarray}
where
$$
C_{\mu} = \left(\sqrt{2} e^{\mu}\right)^{{\cal M}}
\left(\frac{1}{4\cosh{\left(2{\cal N}\omega_{\mu}(0)\right)}
\cosh{\left(2{\cal N}\omega_{\mu}(\pi/2)\right)}}\right)^{\frac{1}{4{\cal N}}}
$$
and a lattice dispersion relation $\omega_{\mu}(x)$  is implicitly given by
$$
\omega_{\mu}(x)={\rm arcsinh}\sqrt{2\sinh^2 \mu+\sin^2 x}
$$
with $\mu = \frac{1}{2} \ln \sinh 2J $.
At the critical point $\mu=\mu_c=0$ ($J_c=\frac{1}{2}\ln{(1+\sqrt{2})}$)
one then obtains
$$
\omega_{0}(x)={\rm arcsinh}\sin x
$$
and
$$
C_0=\left(\sqrt{2}\right)^{{\cal M}}
\left(\frac{1}{4\cosh{\left(2{\cal N}{\rm arcsinh}1\right)}
}\right)^{\frac{1}{4{\cal N}}}
$$
Then the critical free energy $f_{\cal M}$, critical spin-spin
correlation length $\xi_s$ and critical energy-energy correlation
length $\xi_e$ of Eq. (\ref{I61}) can be written as
\begin{eqnarray}
f_{\cal M}&=&\frac{{\cal M}}{2({\cal M}+1)}\ln 2-\frac{1}{2({\cal M}+1)}\ln{(1+\sqrt{2})}+\frac{1}{2({\cal M}+1)}\sum_{m=0}^{2{\cal M} +1}\omega_0
\left(\frac{\pi m}{2({\cal M}+1)}\right),
\label{I81} \\
\xi_e^{-1}&=& 2\omega_0 \left(\frac{\pi}{2({\cal M}+1)}\right).
\label{I82}\\
\xi_s^{-1}&=&
2\omega_0 \left(\frac{\pi}{{\cal M}+1}\right),
\label{I83}
\end{eqnarray}
Using the Euler-Maclaurin summation formula \cite{hardy} the
asymptotic expansion of the critical free energy $f_ {\cal M}$ can
be written in the following form
\begin{eqnarray}
({\cal M}+1)(f_{{\cal M}}- f_{\infty}) &=&2 f_{surf}-
\sum_{k=0}^{\infty}\frac{\lambda_{2k}B_{2 k+2}}{(2k)!(2k+2)}
\left(\frac{\pi}{2({\cal M}+1)}\right)^{2 k+1}
, \label{fN-all1}\\
&=&2 f_{surf}-\frac{\pi}{{24(\cal M}+1)} -\frac{1}{2880} \left(\frac{\pi}{{\cal M}+1}\right)^3-\frac{1}{48384}
 \left(\frac{\pi}{{\cal M}+1}\right)^5
+\dots, \nonumber
\end{eqnarray}
where
\begin{eqnarray}
f_{\infty}&=&\frac{1}{2}\ln{2}+\frac{2 G}{\pi}
\label{fb}\\
f_{surf}&=&-\frac{1}{4}\ln{(2+2\sqrt{2})}
\label{f1}
\end{eqnarray}
and $\lambda_{2k}$ is the coefficients in the
 the Taylor expansion of the $\omega_0(x)$:
\begin{eqnarray}
\omega_0(x)=\sum_{p=0}^{\infty}
\frac{\lambda_{2p}}{(2p)!}\;x^{2p+1}, \qquad \qquad \lambda_0=1,
\lambda_2=-\frac{2}{3}, \lambda_4=4, ...
\label{SpectralFunctionExpansion1}
\end{eqnarray}
Using the Taylor expansion of the $\omega_0(x)$ given by Eq. (\ref{SpectralFunctionExpansion1}) the asymptotic expansion of the critical spin-spin correlation length
$\xi_s$ and critical energy-energy correlation length $\xi_e$ can
be written as
\begin{eqnarray}
\xi_s^{-1} &=& \sum_{k=0}^{\infty}\frac{2\lambda_{2k}}{(2k)!}
\left(\frac{\pi}{2({\cal M}+1)}\right)^{2 k+1},
\label{I191}\\
&=&\frac{\pi}{{\cal M}+1} -\frac{1}{12} \left(\frac{\pi}{{\cal M}+1}\right)^3+\frac{1}{96}
 \left(\frac{\pi}{{\cal M}+1}\right)^5 +\dots,
\nonumber\\
\xi_e^{-1} &=& \sum_{k=0}^{\infty}\frac{2\lambda_{2k}}{(2k)!}
\left(\frac{\pi}{{\cal M}+1}\right)^{2 k+1},
\label{I181}\\
&=&\frac{2 \pi}{{\cal M}+1} -\frac{2}{3} \left(\frac{\pi}{{\cal
M}+1}\right)^3+\frac{1}{3}
 \left(\frac{\pi}{{\cal M}+1}\right)^5
+\dots, \nonumber
\end{eqnarray}
Equations (\ref{fN-all1}), (\ref{I191}), and (\ref{I181}) imply
that the ratios of the amplitudes of the $({\cal M}+1)^{-(2k+1)}$
correction terms in the spin-spin correlation length, the
energy-energy correlation lengths, and the free energy expansion,
{\it i.e.} $b_k/a_k$ and $c_k/a_k$, should not depend in detail on
the dispersion relation ($\omega_{0}(x)$) as given by Eqs. (\ref{new6}) and
(\ref{new7}).

To check the applicability of these results, we study the
anisotropic  Ising model with coupling constant $J$ and $\gamma J$ along
the horizontal and vertical directions, respectively, with $0 < \gamma
< \infty$. At the critical point $J_c$, where $J_c$ is defined by
$\sinh{2J_c}\sinh{2\gamma J_c} = 1$, we obtain that the dispersion relation  for the anisotropic  Ising model $\omega_{0}^{(anysotrop)}(x)$ is now given by
\begin{equation}
\omega_{0}^{(anysotrop)}(x)={\rm arcsinh}\left(\sin
x\;\sinh{2 \gamma J_c}\right)
\label{omegaanysotrop}
\end{equation}
The asymptotic expansion of the critical free energy $f_ {\cal M}$, the critical spin-spin correlation length
$\xi_s$ and critical energy-energy correlation length $\xi_e$ is given by the first line in the Eqs. (\ref{fN-all1}),  (\ref{I191}), and (\ref{I181}), where $\lambda_{2k}$ is now the coefficients in the
 the Taylor expansion of the dispersion relation $\omega_{0}^{(anysotrop)}(x)$:
\begin{eqnarray}
\omega_0^{(anysotrop)}(x)=\sum_{p=0}^{\infty}
\frac{\lambda_{2p}}{(2p)!}\;x^{2p+1},
\label{SpectralFunctionExpansion3}
\end{eqnarray}
where
$$
\lambda_0=\sinh{2 \gamma J_c}, \quad \lambda_2=-\frac{1}{3}\sinh{2
\gamma J_c}\;\cosh^2{2 \gamma J_c}, \quad ...
$$
It is easy to see that the ratios $b_k/a_k$ and $c_k/a_k$ does not
depend in detail on the dispersion relation
($\omega_{0}^{(anysotrop)}(x)$) and Eqs. (\ref{new6}) and
(\ref{new7}) holds for all anisotropy $\gamma$.
%\lambda_4=\frac{1}{5}\sinh{2 \gamma J_c}\cosh^2{2 \gamma
%J_c}(1+9\sinh^2{2 \gamma J_c})

In \cite{izmailan}, we have shown that the Ising model on the square, honeycomb and plan-triangular lattices,
and the quantum spin model have universal amplitude ratios, i.e. we confirmed
that such models are in the same universality class. It is reasonable to expect that the ratios of
Eqs. (\ref{new6}) and (\ref{new7}) are valid for the same set of models with mixed BCs.

The leading terms of Eqs. (\ref{fN-all1}), (\ref{I191}) and
(\ref{I181})  are consistent with Eqs. (\ref{I2}) - (\ref{Afree}),
{\it i.e.} $a_1$, $b_1$ and $c_1$ are universal. Equations
(\ref{I2}) and (\ref{I1}) implies immediately that their ratio is
also universal, namely $r_s(1)=D_1/A$ and $r_e(1)=D_2/A$, which is
consistent with Eqs. (\ref{new6}) and (\ref{new7}) for the case
$k=1$
\begin{eqnarray}
r_s(1)&=&\frac{D_1}{A} =-24 \label{rs1}\\
r_e(1)&=&\frac{D_2}{A} =-48 \label{re1}
\end{eqnarray}

The finite-size corrections to Eqs. (\ref{I2}) and (\ref{I1}) can
be calculated by the means of a perturbated conformal field theory
\cite{cardy86,zamol87}. In general, any lattice Hamiltonian will
contain correction terms to the critical Hamiltonian $H_c$
\begin{equation}
H = H_c + \sum_p g_p \int_{-{\cal M}/2}^{{\cal M}/2}\phi_p(v) d v, \label{Hc}
\end{equation}
where $g_p$ is a non-universal constant and $\phi_p(v)$ is a
perturbative conformal field. Below we will consider the case with
only one perturbative conformal field, say $\phi_l(v)$. Then the
eigenvalues of $H$ are
\begin{equation}
E_n=E_{n,c}+  g_l \int_{-{\cal M}/2}^{{\cal M}/2}<n|\phi_l(v)|n> d v + \dots,
\label{En}
\end{equation}
where $E_{n,c}$ are the critical eigenvalues of $H$. The matrix
element $<n|\phi_l(v)|n>$ can be computed in terms of the
universal structure constants $(C_{nln})$ of the operator product
expansion \cite{cardy86}: $<n|\phi_l(v)|n> =\left({2
\pi}/{{\cal M}}\right)^{x_l}C_{nln}$, where $x_l$ is the scaling
dimension of the conformal field $\phi_l(v)$. The energy gaps
$(E_n-E_0)$ and the ground-state energy ($E_0$) can be
written as
\begin{eqnarray}
E_n-E_0&=&\frac{2 \pi}{{\cal M}} x_n+ 2 \pi
g_l(C_{nln}-C_{0l0})\left(\frac{2 \pi}{{\cal M}}\right)^{x_l-1} + \dots,
\label{xin}\\
E_0 &=& E_{0,c}+2 \pi  g_l C_{0l0} \left(\frac{2
\pi}{{\cal M}}\right)^{x_l-1} + \dots. \label{E0conf}
\end{eqnarray}
Note, that the ground state energy $E_0$, the  first energy gap
($E_1-E_0$) and the second energy gap ($E_2-E_0$) of a quantum
spin chain are, respectively, the quantum analogies of the free
energy $f(N)$, inverse spin-spin correlation length
$\xi_s^{-1}(N)$, and inverse energy-energy correlation length
$\xi_e^{-1}(N)$ for the Ising model; that is,
\begin{equation}
{\cal M} f_{{\cal M}} \Leftrightarrow - E_0, \quad \xi_s^{-1}(N) \Leftrightarrow
E_1-E_0 \equiv \Delta_s, \quad \mbox{and} \quad \xi_e^{-1}(N)
\Leftrightarrow E_2-E_0 \equiv \Delta_e. \label{definition}
\end{equation}
For the 2D Ising model, one finds \cite{henkel} that the leading
finite-size corrections ($1/{\cal M}^3$) can be described by the
Hamiltonian given by Eq. (\ref{Hc}) with a single perturbative
conformal field $\phi_l(v)=L_{-2}^2(v)$ with
scaling dimension $x_l=4$ .

In order to obtain the corrections we need the matrix elements
$<n|L_{-2}^2(v)|n>$, which have already been
computed by Reinicke \cite{reinicke87}.
\begin{eqnarray}
<\Delta+r|L_{-2}^2|\Delta+r> &=&
\left(\frac{2\pi}{{\cal M}}\right)^4\left[\frac{49}{11520}+(\Delta+r)\left(\Delta-\frac{5}{24}+\frac{r(2
\Delta + r)(5 \Delta+1)}{(\Delta+1)(2\Delta+1)}\right)\right]
\label{matrixelement}
\end{eqnarray}
The universal structure constants $C_{2l2}$, $C_{1l1}$ and
$C_{0l0}$ can be obtained from the matrix element
$<n|L_{-2}^2(v)|n> =\left({2
\pi}/{{\cal M}}\right)^{x_l}C_{nln}$, where $x_l = 4$ is the scaling
dimension of the conformal field $L_{-2}^2(v)$.

At the critical point  the spectra of the Hamiltonian with  free,
fixed and mixed BCs are built by the irreducible representation
$\Delta$ of a single Virasoro algebras with possible values of
$\Delta$ are $0, \frac{1}{2}, \frac{1}{16}$. We denote by $\Delta$
the highest weight, and by $\Delta+r$, the $r$-th level having
degeneracy $d(\Delta,r)$ of irreducible representation of the
Virasoro algebra. A state will be labeled by $|n> \sim
|\Delta+r>$.

For mixed BCs the ground state $|0>$, first excited state $|1>$, and second
excited state $|2>$ are given
by \cite{cardy86,gehlen}:
\begin{eqnarray}
|0>&=&|\Delta=\frac{1}{16},r=0>, \label{state0m}\\
|1>&=&|\Delta=\frac{1}{16},r=1>, \label{state1m}\\
|2>&=&|\Delta=\frac{1}{16},r=2>.\label{state2m}
\end{eqnarray}

After reaching this point, one can easily compute the universal
structure constants $C_{2l2}$, $C_{1l1}$ and $C_{0l0}$ for mixed boundary conditions.
The values of $C_{0l0}$, $C_{1l1}$,
$C_{2l2}$ can be obtained from Eqs. (\ref{matrixelement}) -
(\ref{state2m}) and given by:
\begin{equation}
C_{0l0}=-7/1440, \qquad C_{1l1} = 1673/1440, \qquad
C_{2l2}=13433/1440. \label{Cm}
\end{equation}

Equations (\ref{xin}) and (\ref{E0conf}) implies that the ratios
of first-order corrections amplitudes for $E_n-E_0$ ($\xi_n^{-1}$) and $-
E_0$ ($f_{{\cal M}}$) is universal and equal to $(C_{0l0}-C_{nln})/C_{0l0}$, which
is consistent with Eq. (\ref{new6})
for the case $n=1, k=2$
\begin{equation}
r_s(2)=\frac{C_{0l0}-C_{1l1}}{C_{0l0}} = 240 \qquad \mbox{for mixed BCs} \label{rsmixed}
\end{equation}
and with Eq. (\ref{new7}) for the
case $n=2, k=2$
\begin{equation}
r_e(2)=\frac{C_{0l0}-C_{2l2}}{C_{0l0}}=
1920 \qquad \mbox{for mixed BCs} \label{remixed}
\end{equation}

In this paper we present exact calculations for a set of universal
amplitude ratios for the two-dimensional (2D) Ising
models on ${\cal M} \times \infty$ lattice
 for mixed BCs universality class.
  We find that such result are in perfect
agreement with a perturbated conformal field theory scenario
proposed by Cardy \cite{cardy86}.

\section{Acknowledgements}
This work was supported by National Science Council
of the Republic of China (Taiwan) under Grant No. NSC 97-2115-M-001-019-MY3.

%%%%%%%%%%%%%%%%%%%%%%%%%%%%%%%%%%%%%%%%%%%%%%%%%%%%%%%%%%%%%%%%%%%%%%%%%%

%%%%%%%%%%%%%%%%%%%%%%%%%%%%%%%%%%%%%%%%%%%%%%%%%%%%%%%%%%%%%%%%%%%%%%%%%%%

\end{document}